\documentclass[twocolumn,floatfix,prb,a4paper]{revtex4-1}
\usepackage[english]{babel} 
\usepackage[utf8]{inputenc}
\usepackage{color}
\usepackage{natbib}
\usepackage{hyperref}
\usepackage{graphicx}
\usepackage{amsmath}
\usepackage{bm}

\begin{document}
\title{Zero-energy quasiparticles in an interacting nanowire containing a topological Josephson junction} 
\author{Julia Boeyens}
\author{Izak Snyman}
\email{izaksnyman1@gmail.com}
\affiliation{Mandelstam Institute for Theoretical Physics, School of Physics, University of the Witwatersrand, PO Box Wits, Johannesburg, South Africa}
\date{6 March 2020}
\begin{abstract} 
We study a Josephson junction in a Kitaev chain with particle-hole symmetric nearest neighbor interactions. 
When the phase difference across the junction is $\pi$, we show analytically that the full spectrum is fourfold degenerate up to corrections that
vanish exponentially in the system size. The Majorana bound states at the ends of the chain are known to survive interactions. Our result
proves that the same is true for the zero-energy quasiparticle localized at the junction. We further study finite size corrections numerically,
and show how repulsive interactions lead to stronger end-to-end correlations than in a noninteracting system with the same bulk gap. 
\end{abstract}
\maketitle

In many superconducting systems, the fundamental degrees of freedom are noninteracting Bogoliubov quasiparticles. 
However, sometimes interactions cannot be ignored.
For instance, when a nanowire is placed on top of a bulk superconductor, superconductivity is
induced in the wire via the proximity effect, but the superconductor does not necessarily screen 
the short range part of Coulomb interactions intrinsic to the wire.~\cite{Sel2011,Gan2011,Sto2011,Tho2013,Cha2015} 
Due to the 1D nature of the wire, it may not
be possible to treat these interactions by the same mean field techniques
used to account for Cooper pairing in the bulk superconductor.
In this case, the occupation numbers of single-quasiparticle orbitals are not conserved, due to scattering between quasiparticles,
and we are dealing with a non-trivial interacting many-body system.    

For the past decade, systems consisting of a nanowire with strong spin-orbit coupling on top of an s-wave superconductor have been
studied intensively.~\cite{Ore2010} For a suitable choice of system parameters, the wire effectively becomes a spinless 1D p-wave superconductor.
The system hosts a zero-energy quasiparticle orbital comprised of two
spatially separated Majorana bound states localized at the ends of the wire.~\cite{Kit2001,Ali2012,Lei2012,Bee2013} 
 Because filling this orbital costs no energy,
each energy level  of the many-body system is at least twofold degenerate. The ground state degeneracy is topologically protected~\cite{Jer2014,Kla2015,Ali2016,Kaw2017}
and may allow for fault-free processing of quantum information.

However, if the system is not in the ground state, i.e. if it contains finite-energy quasiparticles, and scattering between quasiparticles
cannot be neglected, the occupation number of the zero-energy quasiparticle orbital can change, thus destroying the information
being processed. This is referred to as quasiparticle poisoning.~\cite{Fu2009,Yan2014} 
It is important to ask whether electron-electron interactions in the wire are a significant source of quasiparticle poisoning.
Fendley~\cite{Fen2016} analized a Kitaev chain,~\cite{Kit2001} a minimal model for a spinless p-wave wire, and constructed Majorana end state operators for the interacting
system at the particle-hole symmetric point. The existence of these operators implies that the full spectrum of the interacting system is still twofold degenerate. 
Associated with the degeneracy is a dressed zero-energy quasiparticle mode whose occupation number remains conserved, also when the system
is not in the ground state manifold.
Remarkably then, electron-electron interactions need not lead to quasiparticle poisoning.

When the superconducting phase difference across a Josephson junction in a {\em noninteracting} spinless p-wave wire equals $\pi$,
there is a single Andreev bound state localized around the junction.~\cite{Lut2010}
In this case, the bound state orbital is not associated with spatially separated Majorana operators. 
Nonetheless, it has zero energy and constitutes
an essential ingredient for the $4\pi$ Josephson effect that may serve to detect Majorana bound states.~\cite{Fu2009, Jia2011, Lar2019} 
Its presence in the noninteracting system produces a further twofold degeneracy, resulting in an overall fourfold degeneracy of the spectrum.
Here we ask whether this degeneracy also survives particle-hole symmetric electron-electron interactions. 
In other words, do interactions cause quasiparticle poisoning of the $4\pi$ Josephson effect in a Kitaev chain?
We analytically prove that fourfold degeneracy survives, implying no quasiparticle poisoning, until strong interactions induce a quantum phase transition. 

There is however another way in which interactions do degrade the $4\pi$-Josephson signal, namely by enhancing finite size effects.
A perfect $4\pi$ signal requires the quasiparticle at the junction to have precisely zero energy and an infinite lifetime, 
a condition that is only met in the limit of an infinite wire. In a finite system, there are corrections that 
manifest as a lifting of the fourfold degeneracy. These decay exponentially as a function of the system size,
but may still be non-negligible in realistic systems.~\cite{San2012,Pik2012}
We investigate finite size effects numerically. Not surprisingly, repulsive
electron-electron interactions reduce the bulk gap and enhance finite size corrections. However,
if we compare two wires with the same gap, one with repulsive interactions and the other without,
we still find a significantly slower decay of finite size corrections in the interacting case.
We conclude that repulsive interactions frustrate superconductivity
in a way that cannot be accounted for by a noninteracting model with a reduced gap.

The model we study is based on a Kitaev chain with zero chemical potential, and describes a 1D lattice hosting spinless fermions with hopping $W>0$ and
superconducting pairing $|\Delta|$ between nearest neighbor sites.~\cite{Kit2001} We add short range electron-electron interactions~\cite{Gan2011,Tho2013,Cha2015,Mia2017} 
of strength $U$ and re-adjust the
external potential to maintain particle-hole symmetry in the presence of interactions.
We model a Josephson junction~\cite{Jia2011,Ali2012} by setting superconducting pairing to zero between
a pair of neighboring sites. For sites to the left of the junction, 
pairing terms are of the form $-\Delta(a a_{<} + a_{>} a+ \mbox{h.c.})/2$, where $a$ is the fermion annihilation operator associated with a given site
while $a_{<}$ and $a_{>}$ are respectively associated with its left and right neighbors.
We take $0<\Delta<W$. For a $\pi$ superconducting phase difference across the junction, pairing terms 
to the right of the junction are of the form $-\Delta(a_{<} a + a a_{>}+\mbox{h.c.})/2$. We introduce an index $\alpha\in\{1,2\}$ to distinguish sites on the left and right of the junction,
and an index $j=1,\ldots,N$ to label sites by their distance from the junction. With these conventions, the Hamiltonian that we consider is       
\begin{equation}
H=H_0+H_{\rm int},~H_0=\sum_{\alpha=1}^2 H_\alpha + H_{\rm J},
\end{equation}
where
\begin{align}
H_\alpha=&-\frac{W}{2}\sum_{j=1}^{N-1}\left(a_{\alpha, j}^\dagger a_{\alpha, j+1} +a_{\alpha, j+1}^\dagger a_{\alpha, j}\right)\nonumber\\
&~~~-\frac{\Delta}{2}\sum_{j=1}^{N-1}\left(a_{\alpha, j} a_{\alpha, j+1} +a_{\alpha, j+1}^\dagger a_{\alpha, j}^\dagger\right)
\end{align}
describes the chain to the left ($\alpha=1$) and right ($\alpha=2$) of the Josephson junction.
The coupling between the two sites comprising the junction is given by
\begin{equation}
H_{\rm J}=-\frac{W}{2}\left(a_{1,1}^\dagger a_{2,1}+a_{2,1}^\dagger a_{1,1}\right).
\end{equation}
The interaction term reads
\begin{align}
H_{\rm int}=&U\sum_{\alpha=1}^2\sum_{j=1}^{N-1}\left(n_{\alpha,j}-\frac{1}{2}\right)\left(n_{\alpha,j+1}-\frac{1}{2}\right)\nonumber\\
&+U\left(n_{1,1}-\frac{1}{2}\right)\left(n_{2,1}-\frac{1}{2}\right),
\end{align}
where $n_{\alpha,j}=a_{\alpha,j}^\dagger a_{\alpha,j}$. 
The following should be noted. 
Our proof relies on particle-hole conjugation symmetry together with
properties of the system within a finite distance from the ends, $N$ being assumed sufficiently larger than this distance. Apart from
symmetry constraints, the proof therefore does not depend on the
details near the junction. For instance, it does not matter whether the junction is atomically sharp, whether it sits exactly in the middle of the chain, or how the
junction affects the values of $W$, $U$ and $\Delta$ in its vicinity. We adopt a minimal model here purely in order that we may later numerically calculate
finite size corrections to our analytical results.  

Important for our purposes is that $H$ possesses a particle-hole symmetry, i.e. $PHP^\dagger=H$ where the unitary operator $P$ is defined as
\begin{align}
P=&\gamma_{2,N}^{\rm Re}\gamma_{1,N}^{\rm Re}\gamma_{2,N-1}^{\rm Im}\gamma_{1,N-1}^{\rm Im}\times\ldots\nonumber\\
&\ldots\times\left\{\begin{array}{ll} \gamma_{2,2}^{\rm Re}\gamma_{1,2}^{\rm Re}\gamma_{2,1}^{\rm Im}\gamma_{1,1}^{\rm Im}&\mbox{ if }N\mbox{ is even,}\\
\gamma_{2,2}^{\rm Im}\gamma_{1,2}^{\rm Im}\gamma_{2,1}^{\rm Re}\gamma_{1,1}^{\rm Re}&\mbox{ if }N\mbox{ is odd,}\end{array}\right.
\end{align}
with
\begin{equation}
\gamma_{\alpha,j}^{\rm Re}=a_{\alpha,j}+a_{\alpha,j}^\dagger,~\gamma_{\alpha,j}^{\rm Im}=i\left(a_{\alpha,j}-a_{\alpha,j}^\dagger\right),
\end{equation}
so that 
\begin{equation}
Pa_{\alpha,j}P^\dagger=(-)^{N+\alpha-j-1}a_{\alpha,j}^\dagger.\label{pap}
\end{equation}
Furthermore, we rely on the fact that while $H$ does not conserve fermions, it does conserve fermion parity 
$\Pi=\exp i\pi\sum_{\alpha,j} a_{\alpha,j}^\dagger a_{\alpha,j}$, with $+1=$ even and $-1=$ odd. We note that $P$ also preserves fermion parity.

Fourfold degeneracy will follow if an operator $c_1$ exists such that
\begin{align}
&\Pi c_1 \Pi^\dagger=-c_1,\label{con1}\\
&\{c_1,c_1^\dagger\}=1,~c_1^2=[H,c_1]=0,\label{con2}\\
&Pc_1P^\dagger=c_1^\dagger.\label{con3}
\end{align}
The first condition (\ref{con1}) implies that $c_1$ flips the $a$-fermion parity.
The second set of conditions (\ref{con2}) implies that $c_1$ can be viewed as a fermionic annihilation operator for a zero-energy quasiparticle mode with
occupation number $c_1^\dagger c_1$. The spectrum of $H$ is thus twofold degenerate: for every eigenstate in which the zero-energy mode
is empty, there is an eigenstate with the same energy in which the zero-energy mode is filled, and vice versa. The two states have opposite fermion
parity. Now consider one such pair $\left|E,0,\pm\right>$ and $ \left|E,1,\mp\right>\equiv c_1^\dagger \left|E,0,\pm\right>$. Here the first quantum number
$E$ is energy, the second quantum number ($0$ or $1$) is the occupation number of the zero-energy mode, and the third $\pm$ is the fermion parity.
The third condition (\ref{con3}) implies that there are two more states orthogonal to the above two, that have the same energy.
The first is $\left|E,1,\pm\right>\equiv P\left|E,0,\pm\right>$. Because the Hamiltonian has particle-hole symmetry, it is an eigenstate of 
$H$ with energy $E$. It has parity $\pm$ and hence is orthogonal to $\left|E,1,\mp\right>$. The occupation number of the zero-energy mode is one
because $c_1^\dagger c_1  P\left|E,0,\pm\right>=Pc_1P^\dagger P c_1^\dagger P^\dagger P \left|E,0,\pm\right>=Pc_1c_1^\dagger \left|E,0,\pm\right>=P\left|E,0,\pm\right>$.
Thus it is orthogonal to $\left|E,0,\pm\right>$. The fourth state is straightforwardly seen to be $\left|E,0,\mp\right>=c_1P\left|E,0,\pm\right>$.

In the noninteracting system ($U=0$) 
the operator $c_1$ is built from Majorana end state operators that are zero-energy Bogoliubov quasiparticles (linear combinations of
$a_{\alpha,j}$ and $a_{\alpha,j}^\dagger$). Explicitly
\begin{equation}
\gamma_\alpha^{\rm end}=A_{\rm end}\sum_{n=0}^{\left\lfloor \frac{N}{2} \right\rfloor}\left(-\frac{W-\Delta}{W+\Delta}\right)^n\gamma_{\alpha, N-2n}^{\rm Re},\label{majoranas}
\end{equation}
with $\alpha\in\{1,\,2\}$. $A_{\rm end}$ is a normalization constant that ensures $\left(\gamma_\alpha^{\rm end}\right)^2=1$. We see that the Majorana operators
are exponentially localized, with decay constant
\begin{equation}
\nu=\ln\sqrt{\frac{W+\Delta}{W-\Delta}}.\label{decay}
\end{equation}
It is straightforward to check that $\{\gamma_\alpha^{\rm end},\gamma_\alpha^{\rm end}\}=2\delta_{\alpha,\beta}$ and
\begin{equation}
[H_0,\gamma_{\alpha}^{\rm end}]=0,\label{eq:com}
\end{equation}
up to finite size corrections that vanish like $[(W-\Delta)/(W+\Delta)]^N$. 
We disregard these and similar corrections that vanish exponentially as a function of system size
throughout our analytical work. 
In view of (\ref{eq:com}) we can define a fermion annihilation operator
\begin{equation}
c_1=\gamma_1^{\rm end}+i\gamma_2^{\rm end},\label{c1}
\end{equation}
that satisfies conditions (\ref{con1}) and (\ref{con2}). From the particle-hole conjugation relation (\ref{pap}) for $a$-fermions then follows that
\begin{equation}
P\gamma^{\rm end}_{\alpha}P^\dagger=(-)^{\alpha-1}\gamma_{\alpha}^{\rm end},\label{pendp}
\end{equation}
so that (\ref{con3}) is indeed satisfied. Thus we have rederived the known result that the noninteracting system has fourfold degeneracy, up to
corrections that decay exponentially as a function of system size.~\cite{Lut2010} It was not necessary for our proof to find the Andreev bound state
localized at the junction explicitly. It is however straightforward to do so by 
solving the Bogoliubov de Gennes equation associated with $H_0$. One finds
$c_2=(\gamma_+^{\rm J} + i \gamma_-^{\rm J})/2$ where
\begin{align}
\gamma_\pm^{\rm J}=A_{\rm J}\sum_{n=1}^N &e^{-\nu(n-1)}\left(\sin\frac{\pi n}{2}\pm\cosh\nu\cos\frac{\pi n}{2}\right)\nonumber\\
&\times\frac{1}{\sqrt{2}}\left(\gamma_{1,n}^{\rm Im}\pm \gamma_{2,n}^{\rm Im}\right).
\end{align}
The normalization constant $A_{\rm J}$ is chosen such that $(\gamma_\pm^{\rm J})^2=1$ and $\nu$ is defined as in (\ref{decay}).

It is instructive to ask where the above proof breaks down if there is a zero- rather than $\pi$ phase difference across the junction. 
In that case, pairing on both sides of the junction is of the form $-\Delta(a a_{<} + a_{>} a+ \mbox{h.c.})/2$. In $H_0$, the sign of the pairing
term to the right of the junction must then be reversed, i.e. $\Delta\to-\Delta$ for $\alpha=2$. This has the effect of replacing $\gamma_{2, N-2n}^{\rm Re}$ with
$\gamma_{2, N-2n}^{\rm Im}$ in the expression (\ref{majoranas}) for $\gamma_2^{\rm end}$. As a result, the particle hole conjugation properties
of this right end state operator changes from (\ref{pendp}) to $P \gamma_2^{\rm end}P^\dagger=\gamma_2^{\rm end}$. This brings about the 
key modification. Now instead of (\ref{con3}), $c_1$ becomes invariant under particle-hole conjugation, i.e. $Pc_1P^\dagger=c_1$.
As a consequence $\left|E,0,\pm\right>$ and $P\left|E,0,\pm\right>$ have the same $c_1^\dagger c_1$ eigenvalue, and we can no longer conclude that they are
orthogonal.  

Now we consider the interacting case $H=H_0+H_{\rm int}$. 
It is convenient define Wigner-Jordan spin operators
\begin{align}
&\sigma_{\alpha, j}^x= \gamma_{\alpha,j}^{\rm Re} \prod_{k=j+1}^N\sigma_{\alpha,k}^z,~~\sigma_{\alpha, j}^y=\gamma_{\alpha,j}^{\rm Im}\prod_{k=j+1}^N\sigma_{\alpha,k}^z,\\
&\sigma_{\alpha,j}^{z}=2a_{\alpha j}^\dagger a_{\alpha j}-1.
\end{align}
Operators with the same $\alpha$-index obey angular momentum commutation relations, i.e. $[\sigma_{\alpha,j}^\mu,\sigma_{\alpha,k}^\nu]=2i\delta_{jk}\varepsilon^{\mu\nu\rho}\sigma_{\alpha,k}^\rho$.
In our construction, the commutation relations between operators with
different $\alpha$ are less canonical. For $\mu\in\{x,\,y\,,z\}$ we still have 
$[\sigma_{1j}^\mu,\sigma_{2k}^z]=[\sigma_{1j}^z,\sigma_{2k}^\mu]=0$.
However for $l,\,m\in\{x,\,y\}$, we have
$\{\sigma_{1j}^l,\sigma_{2k}^m\}=0$
instead of the more usual commuting operators. 
While it is obviously possible to alter the definition of the spin operators in such a way that all commutators are canonical, it
turns out that the above construction is more useful for our purposes.
 
Away from the junction ($j>1$), the above Wigner Jordan transformation maps the Kitaev chain onto two spin-1/2 Heisenberg chains with nearest neighbor
exchange couplings 
\begin{equation}
J_x=(W+\Delta)/4,~J_y=(W-\Delta)/4,~J_z=U/4.
\end{equation} 
From the known properties of the XYZ model,~\cite{Jaf2017} it can then be inferred that either attractive or repulsive
interactions close the superconducting gap when $\left|U\right|=W+\Delta$. For interactions stronger than this critical value,
the system is again in a gapped phase.
We collectively refer to the $|U|>W+\Delta$ regimes as the intrinsically gapped phase, because the gap is driven by interactions inside the wire
$\propto U$ rather than by extrinsic pairing $\propto \Delta$.  We refer to the $|U|<W+\Delta$ regime as the
extrinsically gapped phase.  

To prove fourfold degeneracy along the same lines as in the noninteracting case, 
we need to consider the Majorana end state operators of the interacting system.
They have in common with the noninteracting case that
they are Hermitian, decay into the bulk, commute with the Hamiltonian up to terms that vanish exponentially in system size, 
and square to unity. In order to translate the explicit expressions that Fendley~\cite{Fen2016} derived into our notation, we define
\begin{align}
\Psi_{\alpha,xyz}=A_{xyz}\sum_{b=1}^N&\left(\frac{J_y J_z}{J_x^2}\right)^{b-1}
\sigma_{\alpha,N+1-b}^x\nonumber\\
&\times\sum_{s=0}^{\left\lfloor \frac{b-1}{2}\right\rfloor}\sum_{(l_1,\,\ldots,\,l_{2s})}
\prod_{t=1}^{s}Q_{xyz}(\alpha,l_{2t-1},l_{2t}).
\end{align}
Here the sum $\sum_{(l_1,\,\ldots,\,l_{2s})}$ is over all distinct sets of integers $l_1,\,\ldots,\,l_{2s}$ such that
\begin{equation}
N+1-b<l_1<\ldots<l_{2s}<N+1,
\end{equation}
and
\begin{align}
Q_{xyz}(\alpha,j,k)=&\left(\frac{J_z}{J_x}\right)^{j-k}\left(1-\frac{J_x^2}{J_y^2}\right)\sigma_{\alpha,j}^y\sigma_{\alpha,k}^y\nonumber\\
&+\left(\frac{J_y}{J_x}\right)^{j-k}\left(1-\frac{J_x^2}{J_z^2}\right)\sigma_{\alpha,j}^z\sigma_{\alpha,k}^z.
\end{align}
The normalization constant is
\begin{equation}
A_{xyz}=\sqrt{\left(1-\frac{J_y^2}{J_x^2}\right)\left(1-\frac{J_z^2}{J_x^2}\right)}.
\end{equation}
In the extrinsically gapped phase ($|U|<W+\Delta$), the Majorana end state operators are $\gamma_{\alpha,\,{\rm ext}}^{\rm end}=\Psi_{\alpha,xyz}$.
In the intrinsically gapped phase ($|U|>W+\Delta$), the dominant exchange term switches from $x$ to $z$, with the result that the end state operators are obtained from those
of the extrinsically gapped phase by exchanging the roles of $x$ and $z$, i.e. $\gamma_{\alpha,\,{\rm int}}^{\rm end}=\Psi_{\alpha,zyx}$.
A crucial difference between the end state operators in the two phases is immediately apparent: Each term in $\gamma_{\alpha,\,{\rm ext}}^{\rm end}$
contains an unpaired $\sigma_{\alpha,j}^x$ operator and therefore flips fermion number parity, while the corresponding unpaired
operator in the intrinsically gapped phase is $\sigma_{\alpha,j}^z$, which preserve fermion number parity. A related property is that while
$\gamma_{1,\,{\rm ext}}^{\rm end}$ and $\gamma_{2,\,{\rm ext}}^{\rm end}$ anticommute,
$\gamma_{1,\,{\rm int}}^{\rm end}$ and $\gamma_{2,\,{\rm int}}^{\rm end}$ {\em commute}.
The operators also have different behavior under particle-hole conjugation. From the definitions of the spin operators follow that
$P\sigma_{\alpha,j}^{x}P^\dagger=(-)^{\alpha-1}\sigma_{\alpha,j}^{x}$, $P\sigma_{\alpha,j}^{y}P^\dagger=(-)^{\alpha}\sigma_{\alpha,j}^{y}$, and
$P\sigma_{\alpha,j}^{z}P^\dagger=-\sigma_{\alpha,j}^{z}$.
As a result
\begin{equation}
P\gamma_{\alpha,\,{\rm ext}}^{\rm end}P^\dagger=(-)^{\alpha-1}\gamma_{\alpha,\,{\rm ext}}^{\rm end},\label{gammasc}
\end{equation}
as in the noninteracting case, while on the other hand
\begin{equation}
P\gamma_{\alpha,\,{\rm int}}^{\rm end}P^\dagger=-\gamma_{\alpha,\,{\rm int}}^{\rm end}. \label{gammamott}
\end{equation}
In the extrinsically gapped phase we have all the ingredients required to derive fourfold degeneracy. 
Since $\{\gamma_{\alpha,\,{\rm ext}}^{\rm end},\gamma_{\beta,\,{\rm ext}}^{\rm end}\}=2\delta_{\alpha,\beta}$,
we can define a fermionic operator $c_{1,{\rm ext}}=\gamma_{1,{\rm ext}}^{\rm end}+i\gamma_{2,{\rm ext}}^{\rm end}$ in analogy to the noninteracting case
(\ref{c1}). The annihilation operator $c_{1,{\rm ext}}$ is no longer a simple linear combination of $a_{\alpha,j}$
and $a_{\alpha,j}^\dagger$. The interaction dresses the Bogoliubov quasiparticle with particle-hole excitations. Nonetheless,  $c_{1,{\rm ext}}$ meets all three conditions 
(\ref{con1} - \ref{con3}). Thus, in the extrinsically gapped phase, the full spectrum is fourfold degenerate up to corrections that vanish exponentially as a function of the 
length of the wire. Adiabatic continuity with the noninteracting system implies that just as there is a dressed zero-energy fermionic quasiparticle divided between the ends of the system
whose occupation number is conserved, there is one localized around the junction. 
Here we also note that when the phase difference across the junction is $0$ rather than $\pi$, the proof of fourfold degeneracy breaks down in exactly the same way
as in the noninteracting case. In the expression for the right end state operator, the unpaired $\gamma^{\rm Re}_{2,j}$ operator is replaced by a $\gamma^{\rm Im}_{2,j}$,
with the result that the dressed $c_{1,{\rm ext}}$ operator becomes invariant under particle-hole conjugation, rather than transforming into its conjugate, as
is required for fourfold degeneracy.  

In the intrinsically gapped phase, the above construction does not work: because $\gamma_{1,\,{\rm int}}^{\rm end}$ and $\gamma_{2,\,{\rm int}}^{\rm end}$ commute
rather than anticommute, and furthermore preserve fermion number parity, we can no longer construct a fermionic operator in analogy to $c_1$ 
of the noninteracting case (\ref{c1}). It seems the best we can do is to diagonalize $H$, $\gamma_{1,\,{\rm int}}^{\rm end}$ and $\gamma_{2,\,{\rm int}}^{\rm end}$
simultaneously. The eigenvalues $\lambda_1$ and $\lambda_2$ of the latter two operators are $\pm 1$. If $\left|E,\lambda_1,\lambda_2\right>$ is a simultaneous
eigenstate with energy $E$, then $P \left|E,\lambda_1,\lambda_2\right>$ is a different simultaneous eigenstate, that, due to (\ref{gammamott}), has   
$\gamma_{1,\,{\rm int}}^{\rm end}$ and $\gamma_{2,\,{\rm int}}^{\rm end}$ eigenvalues $-\lambda_1$ and $-\lambda_2$. Thus we have only proved
twofold degeneracy. In the limit $|U|\gg W,\,\Delta$, the system becomes equivalent to a spin-1/2 Ising chain, in which case the degeneracy of
the ground state as well as the most excited state is indeed only twofold. 
Below we report a numerical calculation confirming that in the intrinsically gapped phase,
the degeneracy remains twofold as $|U|$ is reduced toward the critical point. 
  
The finite size corrections that we have ignored throughout the above analysis, cause lifting of degeneracies.
While the corrections vanish exponentially in system size, there may be practical considerations limiting how long a real device can be made.
For instance, the longer the wire, the more likely it is to include (rare) regions where the disorder potential is large enough to 
destroy topological order.
Indeed, in Ref.~\onlinecite{Pik2012}, it was estimated that the maximum exponential suppression factor achievable in an InAs wire realization is
of order $10^{-2}$. This estimate did not take interactions into account. It is therefore worth asking how interactions affect finite size corrections.
To address this question, we have performed DMRG calculations on a finite system. Our results are accurate to at least seven significant digits.
We have calculated the ground state and first few excited state energies as a function of $U$, for a system with $\Delta/W=0.2$ and $N=41$ sites
on each side of the junction. (We confined our attention to a few low-lying energy levels because a 
separate and increasingly expensive numerical calculation is required for each higher excited state.)

\begin{figure}[h]
\begin{center}
\includegraphics[width=.99 \columnwidth]{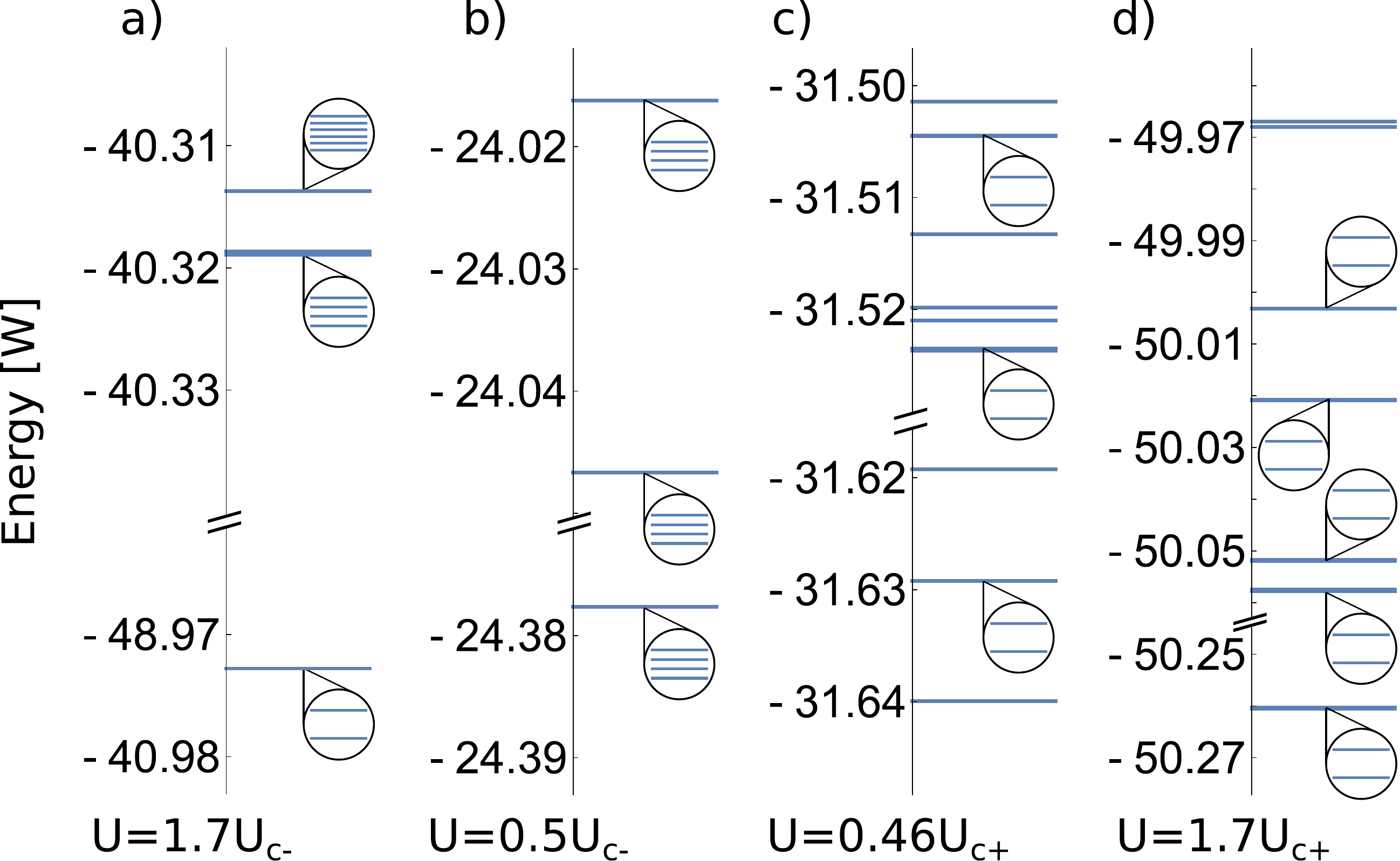}
\end{center}
\caption{Low-lying spectrum for a chain with 41 sites on each side of the Josephson junction and $\pi$-phase difference across the junction, for several 
values of the interaction strength $U$. We used $\Delta/W=0.2$ throughout. Where near-degeneracies cannot be resolved with the naked eye, 
the degeneracy is schematically indicated. 
In panels (a) and (d), the system is in the intrinsically gapped phase, while in panels (b) and (c)
the system is in the extrinsically gapped phase. For attractive interactions [(a) and (b)], finite size errors are too small to be resolved at
the scale of the figure, and the degeneracy structure of the infinite system is clearly seen -- twofold at worst in panel (a) and fourfold in panel (b). For the repulsive interactions
in panel (c), finite size errors are significant and the degeneracy structure of the infinite system's spectrum is obscured. \label{f1}}
\end{figure}

In Figure \ref{f1} we show the low-energy spectrum at a few representative interaction strengths. 
For attractive interactions $U<0$, finite size corrections are small and the 
predicted degeneracy of the infinite system's spectrum is clearly seen:
above the critical point $U_{{\rm c},-}=-(W+\Delta)/2$ (extrinsically gapped phase) there is fourfold degeneracy while below the critical point (intrinsically gapped phase),
some levels, such as the ground state, are only twofold degenerate.  

\begin{figure}[th]
\begin{center}
\includegraphics[width=.99 \columnwidth]{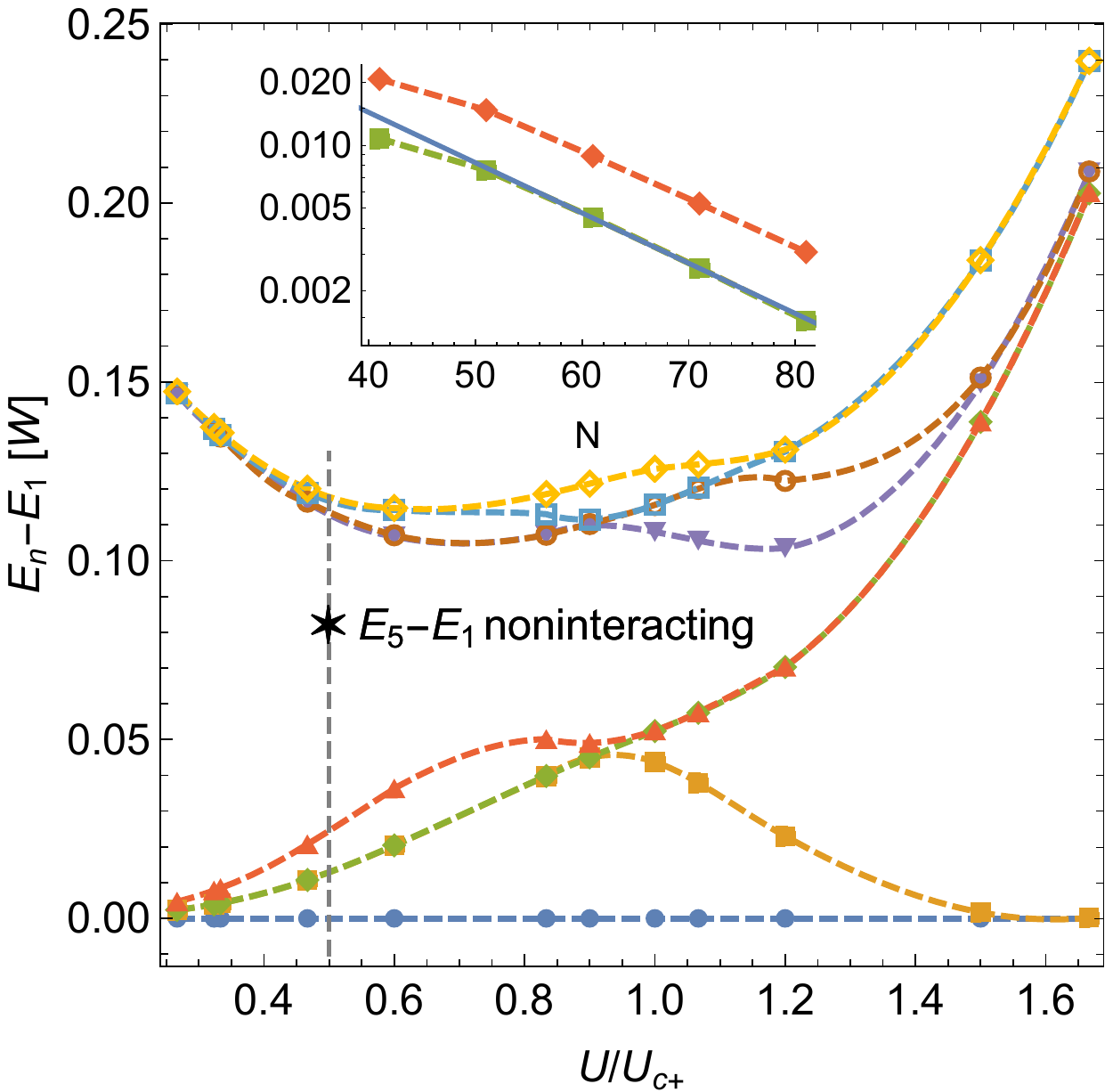}

\end{center}
\caption{Finite size corrections in the case of repulsive interactions. Main panel: Excitation energies $E_n-E_1$, $n=1,\,\ldots 8$, 
for a chain with 41 sites on each side of the Josephson junction and $\pi$-phase difference across the junction,
as a function of $U\in (0.3\, U_{{\rm c}+},1.7\, U_{{\rm c}+})$. $\Delta/W=0.2$ throughout. (Some of this data is also plotted in Figure \ref{f1}). 
Lines are there to guide the eye. The vertical line at $U=0.5 U_{{\rm c}+}$ indicates the point for which we performed a finite size scaling
analysis of $E_n-E_1$, $n=2,\,\ldots 4$. The black star indicates the excitation energy $E_5-E_1$, i.e. from the ground state to the first bulk excitation, in a noninteracting
system with $\Delta=0.054 W$. This is the value of $\Delta$ that produces the same decay of finite size corrections as is observed
in the interacting system with $\Delta=0.2 W$ and $U=0.5 U_{{\rm c}+}$. Inset: Finite size scaling.
The excitation energies $E_n-E_1$, $n=2,\,\ldots 4$, as a function of $N$, for the $U=0.5 U_{{\rm c}+}$ and  $\Delta/W=0.2$. (The $n=2$ and $n=3$ data lie on top of
each other.) The solid curve is the fit $E_2-E_1=0.134\times (0.946)^N$. \label{f2}}
\end{figure}

For repulsive interactions in the window $(0.3\, U_{{\rm c},+},1.5\,U_{{\rm c}+})$ on the other hand, where $U_{{\rm c}+}=W+\Delta$ is the repulsive critical point, 
degeneracy is obscured by finite size effects.
(See main panel of Figure \ref{f2}.)
Whereas on the attractive side, the gap closes linearly in $|U-U_{{\rm c},-}|$, on the repulsive side the gap closes quadratically
in $U-U_{{\rm c},+}$.~\cite{Ale2016} As a result, there is a larger region around the repulsive critical point compared to the attractive critical point where 
the bulk gap is small. It is not surprising that degeneracy lifting in this region should be more noticeable than in regions where
the bulk gap is larger. However, closer inspection of the finite size corrections reveals a non-trivial interaction effect, namely
significantly slower decay than would be the case in a noninteracting system with the same bulk gap. To see this, we fixed
$U=U_{{\rm c}+}/2$ (the vertical line in the main panel of Figure \ref{f2}) and calculated the low-lying excitation energies as a function
of system size. As seen in the inset of Figure \ref{f2}, exponential decay $\sim (0.946)^N$ is observed. We then ask `What value of $\Delta$ would
produce the same decay in a noninteracting system?'. From (\ref{decay}) we get $\Delta=0.054 W$. We also confirmed this numerically (not shown).
In the infinite noninteracting system, this would be the excitation energy between the ground state $E_1$ and the first state above the gap $E_5$, but
to compare with our finite interacting system, we need to take finite size corrections in the noninteracting system into account as well.
In a noninteracting system with 41 sites on each side of the junction and $\Delta=0.054 W$, the excitation energy from the ground state to the first state above the gap is 
$E_5-E1=0.082 W$.
The same excitation energy in the interacting system is $0.116 W$, which is significantly larger. The conclusion is that a given rate of
decay of finite size corrections corresponds to a larger value of the bulk gap  in the system with repulsive interactions than in the noninteracting system. 
Because finite size corrections decay faster for a larger gap, this also means that for a given bulk gap, finite size corrections decay more slowly in
a chain with repulsive interactions than in the noninteracting chain.   

In conclusion, we have shown that the full spectrum of a particle-hole symmetric {\em interacting} Kitaev chain is fourfold degenerate when the system contains a Josephson
junction with a $\pi$-phase difference, up to corrections that vanish exponentially as a function of system size. 
This proves that the occupation number of the dressed Andreev bound state localized around the junction is a constant of motion.
Thus, in a sufficiently long chain, intrinsic electron-electron interactions are not a source quasiparticle poisoning where the $4\pi$ Josephson effect is concerned.
We have numerically studied finite size corrections, and found that they decay more slowly in a chain with repulsive interactions than in a noninteracting chain
with the same gap. This shows that repulsive interactions frustrate superconductivity in a way that cannot be accounted for by an effective noninteracting model.

{\it Acknowledgements --}
This work is based on research supported by the National Research Foundation of South Africa (Grant Number 90657).


\begin{thebibliography}{25}%
\makeatletter
\providecommand \@ifxundefined [1]{%
 \@ifx{#1\undefined}
}%
\providecommand \@ifnum [1]{%
 \ifnum #1\expandafter \@firstoftwo
 \else \expandafter \@secondoftwo
 \fi
}%
\providecommand \@ifx [1]{%
 \ifx #1\expandafter \@firstoftwo
 \else \expandafter \@secondoftwo
 \fi
}%
\providecommand \natexlab [1]{#1}%
\providecommand \enquote  [1]{``#1''}%
\providecommand \bibnamefont  [1]{#1}%
\providecommand \bibfnamefont [1]{#1}%
\providecommand \citenamefont [1]{#1}%
\providecommand \href@noop [0]{\@secondoftwo}%
\providecommand \href [0]{\begingroup \@sanitize@url \@href}%
\providecommand \@href[1]{\@@startlink{#1}\@@href}%
\providecommand \@@href[1]{\endgroup#1\@@endlink}%
\providecommand \@sanitize@url [0]{\catcode `\\12\catcode `\$12\catcode
  `\&12\catcode `\#12\catcode `\^12\catcode `\_12\catcode `\%12\relax}%
\providecommand \@@startlink[1]{}%
\providecommand \@@endlink[0]{}%
\providecommand \url  [0]{\begingroup\@sanitize@url \@url }%
\providecommand \@url [1]{\endgroup\@href {#1}{\urlprefix }}%
\providecommand \urlprefix  [0]{URL }%
\providecommand \Eprint [0]{\href }%
\providecommand \doibase [0]{https://doi.org/}%
\providecommand \selectlanguage [0]{\@gobble}%
\providecommand \bibinfo  [0]{\@secondoftwo}%
\providecommand \bibfield  [0]{\@secondoftwo}%
\providecommand \translation [1]{[#1]}%
\providecommand \BibitemOpen [0]{}%
\providecommand \bibitemStop [0]{}%
\providecommand \bibitemNoStop [0]{.\EOS\space}%
\providecommand \EOS [0]{\spacefactor3000\relax}%
\providecommand \BibitemShut  [1]{\csname bibitem#1\endcsname}%
\let\auto@bib@innerbib\@empty
\bibitem [{\citenamefont {Sela}\ \emph {et~al.}(2011)\citenamefont {Sela},
  \citenamefont {Altland},\ and\ \citenamefont {Rosch}}]{Sel2011}%
  \BibitemOpen
  \bibfield  {author} {\bibinfo {author} {\bibfnamefont {E.}~\bibnamefont
  {Sela}}, \bibinfo {author} {\bibfnamefont {A.}~\bibnamefont {Altland}},\ and\
  \bibinfo {author} {\bibfnamefont {A.}~\bibnamefont {Rosch}},\ }\bibfield
  {title} {\bibinfo {title} {Majorana fermions in strongly interacting helical
  liquids},\ }\href {https://doi.org/10.1103/PhysRevB.84.085114} {\bibfield
  {journal} {\bibinfo  {journal} {Phys. Rev. B}\ }\textbf {\bibinfo {volume}
  {84}},\ \bibinfo {pages} {085114} (\bibinfo {year} {2011})}\BibitemShut
  {NoStop}%
\bibitem [{\citenamefont {Gangadharaiah}\ \emph {et~al.}(2011)\citenamefont
  {Gangadharaiah}, \citenamefont {Braunecker}, \citenamefont {Simon},\ and\
  \citenamefont {Loss}}]{Gan2011}%
  \BibitemOpen
  \bibfield  {author} {\bibinfo {author} {\bibfnamefont {S.}~\bibnamefont
  {Gangadharaiah}}, \bibinfo {author} {\bibfnamefont {B.}~\bibnamefont
  {Braunecker}}, \bibinfo {author} {\bibfnamefont {P.}~\bibnamefont {Simon}},\
  and\ \bibinfo {author} {\bibfnamefont {D.}~\bibnamefont {Loss}},\ }\bibfield
  {title} {\bibinfo {title} {Majorana edge states in interacting
  one-dimensional systems},\ }\href
  {https://doi.org/10.1103/PhysRevLett.107.036801} {\bibfield  {journal}
  {\bibinfo  {journal} {Phys. Rev. Lett.}\ }\textbf {\bibinfo {volume} {107}},\
  \bibinfo {pages} {036801} (\bibinfo {year} {2011})}\BibitemShut {NoStop}%
\bibitem [{\citenamefont {Stoudenmire}\ \emph {et~al.}(2011)\citenamefont
  {Stoudenmire}, \citenamefont {Alicea}, \citenamefont {Starykh},\ and\
  \citenamefont {Fisher}}]{Sto2011}%
  \BibitemOpen
  \bibfield  {author} {\bibinfo {author} {\bibfnamefont {E.~M.}\ \bibnamefont
  {Stoudenmire}}, \bibinfo {author} {\bibfnamefont {J.}~\bibnamefont {Alicea}},
  \bibinfo {author} {\bibfnamefont {O.~A.}\ \bibnamefont {Starykh}},\ and\
  \bibinfo {author} {\bibfnamefont {M.~P.~A.}\ \bibnamefont {Fisher}},\ }\bibfield
   {title} {\bibinfo {title} {Interaction effects in topological
  superconducting wires supporting Majorana fermions},\ }\href
  {https://doi.org/10.1103/PhysRevB.84.014503} {\bibfield  {journal} {\bibinfo
  {journal} {Phys. Rev. B}\ }\textbf {\bibinfo {volume} {84}},\ \bibinfo
  {pages} {014503} (\bibinfo {year} {2011})}\BibitemShut {NoStop}%
\bibitem [{\citenamefont {Thomale}\ \emph {et~al.}(2013)\citenamefont
  {Thomale}, \citenamefont {Rachel},\ and\ \citenamefont
  {Schmitteckert}}]{Tho2013}%
  \BibitemOpen
  \bibfield  {author} {\bibinfo {author} {\bibfnamefont {R.}~\bibnamefont
  {Thomale}}, \bibinfo {author} {\bibfnamefont {S.}~\bibnamefont {Rachel}},\
  and\ \bibinfo {author} {\bibfnamefont {P.}~\bibnamefont {Schmitteckert}},\
  }\bibfield  {title} {\bibinfo {title} {Tunneling spectra simulation of
  interacting Majorana wires},\ }\href
  {https://doi.org/10.1103/PhysRevB.88.161103} {\bibfield  {journal} {\bibinfo
  {journal} {Phys. Rev. B}\ }\textbf {\bibinfo {volume} {88}},\ \bibinfo
  {pages} {161103} (\bibinfo {year} {2013})}\BibitemShut {NoStop}%
\bibitem [{\citenamefont {Chan}\ \emph {et~al.}(2015)\citenamefont {Chan},
  \citenamefont {Chiu},\ and\ \citenamefont {Sun}}]{Cha2015}%
  \BibitemOpen
  \bibfield  {author} {\bibinfo {author} {\bibfnamefont {Y.-H.}\ \bibnamefont
  {Chan}}, \bibinfo {author} {\bibfnamefont {C.-K.}\ \bibnamefont {Chiu}},\
  and\ \bibinfo {author} {\bibfnamefont {K.}~\bibnamefont {Sun}},\ }\bibfield
  {title} {\bibinfo {title} {Multiple signatures of topological transitions for
  interacting fermions in chain lattices},\ }\href
  {https://doi.org/10.1103/PhysRevB.92.104514} {\bibfield  {journal} {\bibinfo
  {journal} {Phys. Rev. B}\ }\textbf {\bibinfo {volume} {92}},\ \bibinfo
  {pages} {104514} (\bibinfo {year} {2015})}\BibitemShut {NoStop}%
\bibitem [{\citenamefont {Oreg}\ \emph {et~al.}(2010)\citenamefont {Oreg},
  \citenamefont {Refael},\ and\ \citenamefont {von Oppen}}]{Ore2010}%
  \BibitemOpen
  \bibfield  {author} {\bibinfo {author} {\bibfnamefont {Y.}~\bibnamefont
  {Oreg}}, \bibinfo {author} {\bibfnamefont {G.}~\bibnamefont {Refael}},\ and\
  \bibinfo {author} {\bibfnamefont {F.}~\bibnamefont {von Oppen}},\ }\bibfield
  {title} {\bibinfo {title} {Helical liquids and Majorana bound states in
  quantum wires},\ }\href {https://doi.org/10.1103/PhysRevLett.105.177002}
  {\bibfield  {journal} {\bibinfo  {journal} {Phys. Rev. Lett.}\ }\textbf
  {\bibinfo {volume} {105}},\ \bibinfo {pages} {177002} (\bibinfo {year}
  {2010})}\BibitemShut {NoStop}%
\bibitem [{\citenamefont {Kitaev}(2001)}]{Kit2001}%
  \BibitemOpen
  \bibfield  {author} {\bibinfo {author} {\bibfnamefont {A.~Y.}\ \bibnamefont
  {Kitaev}},\ }\bibfield  {title} {\bibinfo {title} {Unpaired Majorana fermions
  in quantum wires},\ }\href {https://doi.org/10.1070/1063-7869/44/10s/s29}
  {\bibfield  {journal} {\bibinfo  {journal} {Phys.-Usp.}\ }\textbf
  {\bibinfo {volume} {44}},\ \bibinfo {pages} {131} (\bibinfo {year}
  {2001})}\BibitemShut {NoStop}%
\bibitem [{\citenamefont {Alicea}(2012)}]{Ali2012}%
  \BibitemOpen
  \bibfield  {author} {\bibinfo {author} {\bibfnamefont {J.}~\bibnamefont
  {Alicea}},\ }\bibfield  {title} {\bibinfo {title} {New directions in the
  pursuit of Majorana fermions in solid state systems},\ }\href
  {https://doi.org/10.1088/0034-4885/75/7/076501} {\bibfield  {journal}
  {\bibinfo  {journal} {Reports on Progress in Physics}\ }\textbf {\bibinfo
  {volume} {75}},\ \bibinfo {pages} {076501} (\bibinfo {year}
  {2012})}\BibitemShut {NoStop}%
\bibitem [{\citenamefont {Leijnse}\ and\ \citenamefont
  {Flensberg}(2012)}]{Lei2012}%
  \BibitemOpen
  \bibfield  {author} {\bibinfo {author} {\bibfnamefont {M.}~\bibnamefont
  {Leijnse}}\ and\ \bibinfo {author} {\bibfnamefont {K.}~\bibnamefont
  {Flensberg}},\ }\bibfield  {title} {\bibinfo {title} {Introduction to
  topological superconductivity and Majorana fermions},\ }\href
  {https://doi.org/10.1088/0268-1242/27/12/124003} {\bibfield  {journal}
  {\bibinfo  {journal} {Semiconductor Science and Technology}\ }\textbf
  {\bibinfo {volume} {27}},\ \bibinfo {pages} {124003} (\bibinfo {year}
  {2012})}\BibitemShut {NoStop}%
\bibitem [{\citenamefont {Beenakker}(2013)}]{Bee2013}%
  \BibitemOpen
  \bibfield  {author} {\bibinfo {author} {\bibfnamefont {C.~W.~J.}~\bibnamefont
  {Beenakker}},\ }\bibfield  {title} {\bibinfo {title} {Search for Majorana
  fermions in superconductors},\ }\href
  {https://doi.org/10.1146/annurev-conmatphys-030212-184337} {\bibfield
  {journal} {\bibinfo  {journal} {Annual Review of Condensed Matter Physics}\
  }\textbf {\bibinfo {volume} {4}},\ \bibinfo {pages} {113} (\bibinfo {year}
  {2013})}.\BibitemShut
  {NoStop}%
\bibitem [{\citenamefont {Jermyn}\ \emph {et~al.}(2014)\citenamefont {Jermyn},
  \citenamefont {Mong}, \citenamefont {Alicea},\ and\ \citenamefont
  {Fendley}}]{Jer2014}%
  \BibitemOpen
  \bibfield  {author} {\bibinfo {author} {\bibfnamefont {A.~S.}\ \bibnamefont
  {Jermyn}}, \bibinfo {author} {\bibfnamefont {R.~S.~K.}\ \bibnamefont {Mong}},
  \bibinfo {author} {\bibfnamefont {J.}~\bibnamefont {Alicea}},\ and\ \bibinfo
  {author} {\bibfnamefont {P.}~\bibnamefont {Fendley}},\ }\bibfield  {title}
  {\bibinfo {title} {Stability of zero modes in parafermion chains},\ }\href
  {https://doi.org/10.1103/PhysRevB.90.165106} {\bibfield  {journal} {\bibinfo
  {journal} {Phys. Rev. B}\ }\textbf {\bibinfo {volume} {90}},\ \bibinfo
  {pages} {165106} (\bibinfo {year} {2014})}\BibitemShut {NoStop}%
\bibitem [{\citenamefont {Klassen}\ and\ \citenamefont {Wen}(2015)}]{Kla2015}%
  \BibitemOpen
  \bibfield  {author} {\bibinfo {author} {\bibfnamefont {J.}~\bibnamefont
  {Klassen}}\ and\ \bibinfo {author} {\bibfnamefont {X.-G.}\ \bibnamefont
  {Wen}},\ }\bibfield  {title} {\bibinfo {title} {Topological degeneracy
  (Majorana zero-mode) and 1+1d fermionic topological order in a magnetic chain
  on superconductor via spontaneous $Z_2^f$ symmetry breaking},\ }\href
  {https://doi.org/10.1088/0953-8984/27/40/405601} {\bibfield  {journal}
  {\bibinfo  {journal} {Journal of Physics: Condensed Matter}\ }\textbf
  {\bibinfo {volume} {27}},\ \bibinfo {pages} {405601} (\bibinfo {year}
  {2015})}\BibitemShut {NoStop}%
\bibitem [{\citenamefont {Alicea}\ and\ \citenamefont
  {Fendley}(2016)}]{Ali2016}%
  \BibitemOpen
  \bibfield  {author} {\bibinfo {author} {\bibfnamefont {J.}~\bibnamefont
  {Alicea}}\ and\ \bibinfo {author} {\bibfnamefont {P.}~\bibnamefont
  {Fendley}},\ }\bibfield  {title} {\bibinfo {title} {Topological phases with
  parafermions: Theory and blueprints},\ }\href
  {https://doi.org/10.1146/annurev-conmatphys-031115-011336} {\bibfield
  {journal} {\bibinfo  {journal} {Annual Review of Condensed Matter Physics}\
  }\textbf {\bibinfo {volume} {7}},\ \bibinfo {pages} {119} (\bibinfo {year}
  {2016})}. \BibitemShut
  {NoStop}%
\bibitem [{\citenamefont {Kawabata}\ \emph {et~al.}(2017)\citenamefont
  {Kawabata}, \citenamefont {Kobayashi}, \citenamefont {Wu},\ and\
  \citenamefont {Katsura}}]{Kaw2017}%
  \BibitemOpen
  \bibfield  {author} {\bibinfo {author} {\bibfnamefont {K.}~\bibnamefont
  {Kawabata}}, \bibinfo {author} {\bibfnamefont {R.}~\bibnamefont {Kobayashi}},
  \bibinfo {author} {\bibfnamefont {N.}~\bibnamefont {Wu}},\ and\ \bibinfo
  {author} {\bibfnamefont {H.}~\bibnamefont {Katsura}},\ }\bibfield  {title}
  {\bibinfo {title} {Exact zero modes in twisted Kitaev chains},\ }\href
  {https://doi.org/10.1103/PhysRevB.95.195140} {\bibfield  {journal} {\bibinfo
  {journal} {Phys. Rev. B}\ }\textbf {\bibinfo {volume} {95}},\ \bibinfo
  {pages} {195140} (\bibinfo {year} {2017})}\BibitemShut {NoStop}%
\bibitem [{\citenamefont {Fu}\ and\ \citenamefont {Kane}(2009)}]{Fu2009}%
  \BibitemOpen
  \bibfield  {author} {\bibinfo {author} {\bibfnamefont {L.}~\bibnamefont
  {Fu}}\ and\ \bibinfo {author} {\bibfnamefont {C.~L.}\ \bibnamefont {Kane}},\
  }\bibfield  {title} {\bibinfo {title} {Josephson current and noise at a
  superconductor/quantum-spin-Hall-insulator/superconductor junction},\ }\href
  {https://doi.org/10.1103/PhysRevB.79.161408} {\bibfield  {journal} {\bibinfo
  {journal} {Phys. Rev. B}\ }\textbf {\bibinfo {volume} {79}},\ \bibinfo
  {pages} {161408} (\bibinfo {year} {2009})}\BibitemShut {NoStop}%
\bibitem [{\citenamefont {Yang}\ and\ \citenamefont {Feldman}(2014)}]{Yan2014}%
  \BibitemOpen
  \bibfield  {author} {\bibinfo {author} {\bibfnamefont {G.}~\bibnamefont
  {Yang}}\ and\ \bibinfo {author} {\bibfnamefont {D.~E.}\ \bibnamefont
  {Feldman}},\ }\bibfield  {title} {\bibinfo {title} {Exact zero modes and
  decoherence in systems of interacting Majorana fermions},\ }\href
  {https://doi.org/10.1103/PhysRevB.89.035136} {\bibfield  {journal} {\bibinfo
  {journal} {Phys. Rev. B}\ }\textbf {\bibinfo {volume} {89}},\ \bibinfo
  {pages} {035136} (\bibinfo {year} {2014})}\BibitemShut {NoStop}%
\bibitem [{\citenamefont {Fendley}(2016)}]{Fen2016}%
  \BibitemOpen
  \bibfield  {author} {\bibinfo {author} {\bibfnamefont {P.}~\bibnamefont
  {Fendley}},\ }\bibfield  {title} {\bibinfo {title} {Strong zero modes and
  eigenstate phase transitions in the {XYZ}/interacting Majorana chain},\
  }\href {https://doi.org/10.1088/1751-8113/49/30/30lt01} {\bibfield  {journal}
  {\bibinfo  {journal} {Journal of Physics A: Mathematical and Theoretical}\
  }\textbf {\bibinfo {volume} {49}},\ \bibinfo {pages} {30LT01} (\bibinfo
  {year} {2016})}\BibitemShut {NoStop}%
\bibitem [{\citenamefont {Lutchyn}\ \emph {et~al.}(2010)\citenamefont
  {Lutchyn}, \citenamefont {Sau},\ and\ \citenamefont {Das~Sarma}}]{Lut2010}%
  \BibitemOpen
  \bibfield  {author} {\bibinfo {author} {\bibfnamefont {R.~M.}\ \bibnamefont
  {Lutchyn}}, \bibinfo {author} {\bibfnamefont {J.~D.}\ \bibnamefont {Sau}},\
  and\ \bibinfo {author} {\bibfnamefont {S.}~\bibnamefont {Das~Sarma}},\
  }\bibfield  {title} {\bibinfo {title} {Majorana fermions and a topological
  phase transition in semiconductor-superconductor heterostructures},\ }\href
  {https://doi.org/10.1103/PhysRevLett.105.077001} {\bibfield  {journal}
  {\bibinfo  {journal} {Phys. Rev. Lett.}\ }\textbf {\bibinfo {volume} {105}},\
  \bibinfo {pages} {077001} (\bibinfo {year} {2010})}\BibitemShut {NoStop}%
\bibitem [{\citenamefont {Jiang}\ \emph {et~al.}(2011)\citenamefont {Jiang},
  \citenamefont {Pekker}, \citenamefont {Alicea}, \citenamefont {Refael},
  \citenamefont {Oreg},\ and\ \citenamefont {von Oppen}}]{Jia2011}%
  \BibitemOpen
  \bibfield  {author} {\bibinfo {author} {\bibfnamefont {L.}~\bibnamefont
  {Jiang}}, \bibinfo {author} {\bibfnamefont {D.}~\bibnamefont {Pekker}},
  \bibinfo {author} {\bibfnamefont {J.}~\bibnamefont {Alicea}}, \bibinfo
  {author} {\bibfnamefont {G.}~\bibnamefont {Refael}}, \bibinfo {author}
  {\bibfnamefont {Y.}~\bibnamefont {Oreg}},\ and\ \bibinfo {author}
  {\bibfnamefont {F.}~\bibnamefont {von Oppen}},\ }\bibfield  {title} {\bibinfo
  {title} {Unconventional Josephson signatures of Majorana bound states},\
  }\href {https://doi.org/10.1103/PhysRevLett.107.236401} {\bibfield  {journal}
  {\bibinfo  {journal} {Phys. Rev. Lett.}\ }\textbf {\bibinfo {volume} {107}},\
  \bibinfo {pages} {236401} (\bibinfo {year} {2011})}\BibitemShut {NoStop}%
\bibitem [{\citenamefont {Laroche}\ \emph {et~al.}(2019)\citenamefont
  {Laroche}, \citenamefont {Bouman}, \citenamefont {van Woerkom}, \citenamefont
  {Proutski}, \citenamefont {Murthy}, \citenamefont {Pikulin}, \citenamefont
  {Nayak}, \citenamefont {van Gulik}, \citenamefont {Nyg{\aa}rd}, \citenamefont
  {Krogstrup}, \citenamefont {Kouwenhoven},\ and\ \citenamefont
  {Geresdi}}]{Lar2019}%
  \BibitemOpen
  \bibfield  {author} {\bibinfo {author} {\bibfnamefont {D.}~\bibnamefont
  {Laroche}}, \bibinfo {author} {\bibfnamefont {D.}~\bibnamefont {Bouman}},
  \bibinfo {author} {\bibfnamefont {D.~J.}\ \bibnamefont {van Woerkom}},
  \bibinfo {author} {\bibfnamefont {A.}~\bibnamefont {Proutski}}, \bibinfo
  {author} {\bibfnamefont {C.}~\bibnamefont {Murthy}}, \bibinfo {author}
  {\bibfnamefont {D.~I.}\ \bibnamefont {Pikulin}}, \bibinfo {author}
  {\bibfnamefont {C.}~\bibnamefont {Nayak}}, \bibinfo {author} {\bibfnamefont
  {R.~J.~J.}\ \bibnamefont {van Gulik}}, \bibinfo {author} {\bibfnamefont
  {J.}~\bibnamefont {Nyg{\aa}rd}}, \bibinfo {author} {\bibfnamefont
  {P.}~\bibnamefont {Krogstrup}}, \bibinfo {author} {\bibfnamefont {L.~P.}\
  \bibnamefont {Kouwenhoven}},\ and\ \bibinfo {author} {\bibfnamefont
  {A.}~\bibnamefont {Geresdi}},\ }\bibfield  {title} {\bibinfo {title}
  {Observation of the 4$\pi$-periodic Josephson effect in indium arsenide
  nanowires},\ }\href {https://doi.org/10.1038/s41467-018-08161-2} {\bibfield
  {journal} {\bibinfo  {journal} {Nature Communications}\ }\textbf {\bibinfo
  {volume} {10}},\ \bibinfo {pages} {245} (\bibinfo {year} {2019})}\BibitemShut
  {NoStop}%
\bibitem [{\citenamefont {San-Jose}\ \emph {et~al.}(2012)\citenamefont
  {San-Jose}, \citenamefont {Prada},\ and\ \citenamefont {Aguado}}]{San2012}%
  \BibitemOpen
  \bibfield  {author} {\bibinfo {author} {\bibfnamefont {P.}~\bibnamefont
  {San-Jose}}, \bibinfo {author} {\bibfnamefont {E.}~\bibnamefont {Prada}},\
  and\ \bibinfo {author} {\bibfnamefont {R.}~\bibnamefont {Aguado}},\
  }\bibfield  {title} {\bibinfo {title} {ac Josephson effect in finite-length
  nanowire junctions with Majorana modes},\ }\href
  {https://doi.org/10.1103/PhysRevLett.108.257001} {\bibfield  {journal}
  {\bibinfo  {journal} {Phys. Rev. Lett.}\ }\textbf {\bibinfo {volume} {108}},\
  \bibinfo {pages} {257001} (\bibinfo {year} {2012})}\BibitemShut {NoStop}%
\bibitem [{\citenamefont {Pikulin}\ and\ \citenamefont
  {Nazarov}(2012)}]{Pik2012}%
  \BibitemOpen
  \bibfield  {author} {\bibinfo {author} {\bibfnamefont {D.~I.}\ \bibnamefont
  {Pikulin}}\ and\ \bibinfo {author} {\bibfnamefont {Yu.~V.}\ \bibnamefont
  {Nazarov}},\ }\bibfield  {title} {\bibinfo {title} {Phenomenology and
  dynamics of a Majorana Josephson junction},\ }\href
  {https://doi.org/10.1103/PhysRevB.86.140504} {\bibfield  {journal} {\bibinfo
  {journal} {Phys. Rev. B}\ }\textbf {\bibinfo {volume} {86}},\ \bibinfo
  {pages} {140504} (\bibinfo {year} {2012})}\BibitemShut {NoStop}%
\bibitem [{\citenamefont {Miao}\ \emph {et~al.}(2017)\citenamefont {Miao},
  \citenamefont {Jin}, \citenamefont {Zhang},\ and\ \citenamefont
  {Zhou}}]{Mia2017}%
  \BibitemOpen
  \bibfield  {author} {\bibinfo {author} {\bibfnamefont {J.-J.}\ \bibnamefont
  {Miao}}, \bibinfo {author} {\bibfnamefont {H.-K.}\ \bibnamefont {Jin}},
  \bibinfo {author} {\bibfnamefont {F.-C.}\ \bibnamefont {Zhang}},\ and\
  \bibinfo {author} {\bibfnamefont {Y.}~\bibnamefont {Zhou}},\ }\bibfield
  {title} {\bibinfo {title} {Exact solution for the interacting Kitaev chain at
  the symmetric point},\ }\href
  {https://doi.org/10.1103/PhysRevLett.118.267701} {\bibfield  {journal}
  {\bibinfo  {journal} {Phys. Rev. Lett.}\ }\textbf {\bibinfo {volume} {118}},\
  \bibinfo {pages} {267701} (\bibinfo {year} {2017})}\BibitemShut {NoStop}%
\bibitem [{\citenamefont {Jafari}(2017)}]{Jaf2017}%
  \BibitemOpen
  \bibfield  {author} {\bibinfo {author} {\bibfnamefont {S.~A.}\ \bibnamefont
  {Jafari}},\ }\bibfield  {title} {\bibinfo {title} {Exact phase boundaries and
  topological phase transitions of the $XYZ$ spin chain},\ }\href
  {https://doi.org/10.1103/PhysRevE.96.012159} {\bibfield  {journal} {\bibinfo
  {journal} {Phys. Rev. E}\ }\textbf {\bibinfo {volume} {96}},\ \bibinfo
  {pages} {012159} (\bibinfo {year} {2017})}\BibitemShut {NoStop}%
\bibitem [{\citenamefont {Alexandradinata}\ \emph {et~al.}(2016)\citenamefont
  {Alexandradinata}, \citenamefont {Regnault}, \citenamefont {Fang},
  \citenamefont {Gilbert},\ and\ \citenamefont {Bernevig}}]{Ale2016}%
  \BibitemOpen
  \bibfield  {author} {\bibinfo {author} {\bibfnamefont {A.}~\bibnamefont
  {Alexandradinata}}, \bibinfo {author} {\bibfnamefont {N.}~\bibnamefont
  {Regnault}}, \bibinfo {author} {\bibfnamefont {C.}~\bibnamefont {Fang}},
  \bibinfo {author} {\bibfnamefont {M.~J.}\ \bibnamefont {Gilbert}},\ and\
  \bibinfo {author} {\bibfnamefont {B.~A.}\ \bibnamefont {Bernevig}},\
  }\bibfield  {title} {\bibinfo {title} {Parafermionic phases with symmetry
  breaking and topological order},\ }\href
  {https://doi.org/10.1103/PhysRevB.94.125103} {\bibfield  {journal} {\bibinfo
  {journal} {Phys. Rev. B}\ }\textbf {\bibinfo {volume} {94}},\ \bibinfo
  {pages} {125103} (\bibinfo {year} {2016})}\BibitemShut {NoStop}%
\end{thebibliography}
\end{document}